\documentclass[twocolumn,showpacs,preprintnumbers,amsmath,amssymb]{revtex4}

\usepackage{graphicx}
\usepackage{dcolumn}
\usepackage{bm}

\begin{document}


\title{Empirical Formulae of Electrical and Thermal Conductivities of Elemental Metals at Room Temperature Ranges\\}

\author{Tadashi  Hirayama}
\affiliation{National Astronomical Observatory of Japan, Mitaka, Tokyo, Japan 118-8588}%


\date{\today}

\begin{abstract}
We propose empirical formulae of the electrical conductivity $\sigma$
and thermal conductivity $\lambda$ for elemental metals such as Na, Cu or Fe
at room temperature ranges. Assuming the relaxation time $\tau=\hbar/k_{\rm B}T$ for all metals,
we propose $\sigma=e^{2}n_{\rm atom}\tau/(mG)$ ($m$=electron mass; $n_{\rm atom}$=number density of atoms in each metal).
If we adopt that a single free parameter $G$ is the sum of outer electron numbers in electron
configuration such as $G=1$ for Cu(4s$^{1})$, $G=\!1+2\!=\!3$ for In$^{49}$(5s$^{2}$4p$^{1}$)
 and $G=\!5$ for Nb$^{41}$(3d$^{4}$4s$^{1}$), the `absolute values' of
$\sigma$ and a similar one for $\lambda$ agree with experiments
within $\sim 20{\%}$ for the majority of metals even including semimetals. 
\end{abstract}
\pacs {72.15.Eb, 72.15.Lh}
\maketitle               
$\textit{Proposed Formulae.}$-- In this paper we treat only elemental metals 
 such as Na, Cu, Fe, etc. without impurity and at 1 atm in the room temperature
 range. Conventional formulae \cite{aschcroft,ibach,kittel} for the electrical
 conductivity $\sigma$ and thermal conductivity $\lambda$ in the free
 electron model are $\sigma=(\tau n_{\rm e}/{m^\ast})e^{2}$ and
 $\lambda=(\tau n_{\rm e}/m^{\ast})\pi^{2}k_{\rm B}^{2}T/3$, respectively. Here $\tau$ is the relaxation time of electrons at the Fermi energy $E_{F}$, $m^{\ast}$ is the
 effective electron mass and $n_{\rm e}$ is the electron number density of metals,
 satisfying $n_{\rm e}=Zn_{\rm atom}$, where $n_{\rm atom}$=metal density/atom weight [$\rm m^{-3}$].
 In order to derive `the absolute values' of
 $\sigma$ and $\lambda$ one needs to know $\tau$, $Z$ and $m^{\ast}$,
 all of which are poorly known for many metals, especially $\tau$ (if
 not from the observations). 

For example, we note that $\tau\sim \hbar/k_{\rm B}T$  was 
claimed for $T\!\gg\!\it \Theta \rm$ \cite{lifshitz,abrikosov}  
and Abrikosov \cite{abrikosov} extends to use it also for $T\approx \it\Theta$  as in eq. (4.18) (see foot note therein). 
Here $\it\Theta$ is the Debye temperature. The accuracy of the `tilde' signs they used is, however, not clear, but also whether
it can be used other than monovalent metals is not clearly stated. 
Pippard \cite{pippard} holds a high opinion of the 1937-Bardeen \cite{bardeen} calculation for monovalent metals (deformed potential $C$) on the absolute $\sigma$ value of Na and K, 
while Ziman \cite{ziman} expressed that it is not very accurate, probably because of a factor of two to three difference 
between experiments of $\sigma_{obs}$
and the Bardeen theory for Rb, Cs, Cu, Ag, and Au, where $\sigma \propto$ ($E_{\rm F}/C)^{2}$.
A concise derivation of the relaxation time $\tau\equiv 1/W$ is in Kittel's 8th edition, Appendix J \cite{kittel} with a
slightly modified description from its 7th edition, which still needs the values of $C, m^{*}/m$ and $c_{\rm s}$
(the sound speed).
Here again no statements were made on non monovalent metals. 
Aschcroft and Mermin, on the other hand, in foot note 7 of Chap. 1 \cite{aschcroft} cast doubt even computations on 
individual particle collisions. 

 In this paper we give these absolute
 values which are in good accord with the observations by adopting
 assumptions below.
We assume for `all' elemental metals
\begin{equation}
\tau =\hbar/k_{\rm B}T,\label{eq.1}
\end{equation}
and introduce a
non-dimensional parameter $G$ in place of the conventional
$m^{\ast}/(mZ)$ appearing in the Drude formula. We then propose
\begin{equation} 
\left(\begin{array}{c}\sigma\\\lambda \end{array}\right)
=\frac{e^{2}n_{\rm atom}\tau}{m}\frac{1}{G}\left(\begin{array}{c}1\\ \pi^{2}k_{\rm B}^{2}T/(3e^{2}) \end{array}\right).
\label{eq.2}
\end{equation} 
Here $\sigma$ is in $\rm\Omega^{-1}m^{-1}$ and $\lambda$ in $\rm Wm^{-1}K^{-1}$ unit, $\hbar=h/2\pi (h$ is the Planck constant), and $k_{B}$ is the Boltzmann constant.

These two equations, eqs.(1) and (2) `combined', do not seem to have been proposed in the past. Certainly, there have been attempts
to relate $\sigma$ to electronic configuration \cite{gerritsen} as early as 1956, but combination of parameters, such as $G$
and $n_{\rm atom}$, are unlike the present one, namely effectively unsuccessful, otherwise usual textbooks could have presented in short
sentences as in our abstract.

Note that eq. (\ref{eq.2}) gives $\sigma \propto 1/T$ due to $\tau\propto 1/T$ and temperature
`independent'-$\lambda$ both being consistent with observations in the
room temperature range (`independent' means as compared to $\sigma
\propto 1/T$ variation). Since $n_{\rm atom}$ for each metal (from density and atomic weight) and $T$ can be given, the
only non-dimensional parameter is $G$. Even if $\tau$ is different from eq. (\ref{eq.1}), unknown departure factor from it  
can be included in $G$ (e.g. $m^{\ast}/m$ or deformed potential in a non-dimensional constant).
Thus assuming $\tau=\hbar/k_{\rm B}T$, we first empirically determine the parameter $G_{\rm obs}$ using observed $\sigma_{\rm obs}$ for each metal. Then we assign
$G$(guessed)-values to be the sum of the outer electron numbers in
electron configuration which are `close' to $G_{\rm obs}$. 

\begin{figure}
\begin{center}
\includegraphics[width=8.5cm]{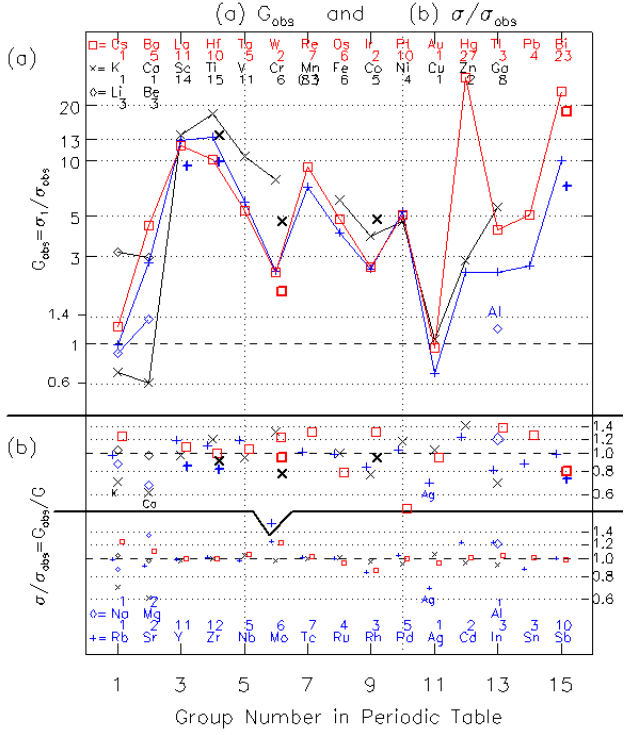}
\end{center}
\caption{(a) $G_{\rm obs} \equiv \sigma_{1}/ \sigma_{\rm obs}$ and (b) $\sigma /
\sigma_{\rm obs}=G_{\rm obs}/G$ against group number for each period, both in
the same logarithmic scales. Thick marks in (a) and (b)-upper are
 $\lambda_{1} / \lambda_{\rm obs}$ values for 6 metals with
 $\vert C_{\rm WF}-1 \vert > 0.2$ (see text). Horizontal positions for some
 metals are slightly shifted to avoid overlapping. The figures in the
 top and bottom are adopted $G$-values.}
\label{f1}
\end{figure}

\textit{Comparison with Experiments.}--
Fig. 1(a) presents $G_{\rm obs}$ plotted against `group' number for each
`period' in the periodic table. Here $G_{\rm obs}$ is defined as
\begin{equation}
G_{\rm obs}\equiv \frac{\sigma_{1}}{\sigma_{\rm obs}}\;\,\,\mathrm{and}\quad 
 \sigma_{1} \equiv \sigma(G=1)=\frac{e^{2}n_{\rm atom}}{m}\frac{\hbar}{k_{\rm B}T}. \label{eq.3}
\end{equation}
Then $G_{\rm obs}$ can be given for each metal from $\sigma_{\rm obs},n_{\rm atom}$ and
temperature $T$ used in the observations. Observed values
($\sigma_{\rm obs}$ and $\lambda_{\rm obs})$ for 48 metals are taken from
Kittel\cite{kittel}, adding $\lambda$(Ca)=201Wm$^{-1}$K$^{-1}$ from the table of Phys. Soc. Japan
\cite{psj} (PSJ-table). We adopt $T_{\rm obs}$=295K from the
Kittel's tabulation for $\sigma_{\rm obs}$. This gives $\tau=2.59\times 10^{-14}$s from eq.(\ref{eq.1}),
which is very close to $\tau_{\rm obs}$ from the observed $\sigma_{\rm obs}$ such as 
$\tau_{\rm obs}$(Na)=2.9$\times 10^{-14}$s and $\tau_{\rm obs}$(Cu)=2.5$\times10^{-14}$s,
using eq.(2) for $G=1$.

First, we find similar trends of $G_{\rm obs}$ among periods of 4
(K$^{19}\sim $Ga$^{31}$), 5 (Rb$^{37}\sim $Sb$^{51}$) and 6
(Cs$^{55}\sim $Bi$^{83}$, inclusive of La$^{57}$). This suggests that
electron configuration, which is the basis of the periodic table, may be
responsible [see e.g. early attempts in figs. 12-13 of ref.\cite{gerritsen}]. Secondly, $G_{\rm obs} \approx 1$ is found for Na$^{11}$,
K$^{19}$, Rb$^{37}$, Cs$^{55}$ and noble metals of Cu$^{29}$, Ag$^{47}$
and Au$^{79}$. This indicates that eq. (\ref{eq.2}) for the electrical
conductivity $\sigma$ with $G=1$ agrees with the observations without
further parameters (Fig.1(b)-upper). 
These 7 elements in free atomic form are in $n$s$^{1}$ outermost electron,
where $n$ is the principal quantum number of $3 \sim 6$.
Though one might say that $Z=1$ and $m^{\ast}=m$ hold as 
expected, $\tau$ should be specified as we propose. 

Third, many metals appear concentrated in $G_{\rm obs}=1$, 3, 5 and $10\sim  
13$, which suggests discreteness of $1/\sigma_{\rm obs}$ if expressed in
unit of $1/\sigma_{1}$, namely $G$(guessed) may well be quantized!
Further we find that in the Kittel's periodic table (K-P-table; in the back cover of the text), not
necessarily in other authors' tables, sum of numbers in the outer
electronic configuration matches the observed $G_{\rm obs}$ quite well. In
fact we find that besides $G \approx 1$ (from $G_{\rm obs} \approx 1$) for
s$^{1}$-electron atoms, $G=3$ (from $G_{\rm obs} \approx 3$) for
3-outer-electron atoms as in In$^{49}$(5s$^{2}$5p$^{1})$, and $G=5$ for
5-outer-electron atoms as in Nb$^{41}$(4d$^{4}$5s$^{1})$. Other examples
besides $G=1$, 3 and 5 are Mg$^{12}$(3s$^{2} \rightarrow G=2$),
Cr$^{24}$(3d$^{5}$4s$^{1}\rightarrow G=6$), Zn$^{30}$(4s$^{2}$
$\rightarrow G=2$), Tc$^{43}$(4d$^{5}$5s$^{2}\rightarrow G=7$; not 4d$^{6}$5s$^{1}$), and
Pb$^{82}$(6s$^{2}$6p$^{2}\rightarrow G=4$). These estimated $G$-values
are shown in Fig. \ref{f1} in the top and bottom. While the K-P-table
actually lists as 3d$^{10}$4s$^{2}$ for Zn, we ignore 3d$^{10}$, as we
ignore the same $n$d$^{10}$($n=3\sim 5$) in Cu, Ag, Au, and Cd$^{48}$(5s$^{2} \rightarrow G$=2) ; adding 10 to $G$ is far
 beyond the observation, though $n$d$^{10}$ may be important for the electronic structure.  In the case of Cu, we know that 
the state density from d$^{10}$-orbits 
is confined below the Fermi energy and hence no contribution (see Fig. 7.12 of ref. 2).
Adding further Al$^{13}$($G$=1, only 2p$^{1}$ is used, since $G_{\rm obs}$=1.2), Sr$^{38}$, Ta$^{73}$, W$^{74}$($G$=2, 6s$^{2}$), Re$^{75}$, Ir$^{77}$($G$=2, 6s$^{2}$), and Tl$^{81}$, 
altogether 22 metals in the K-P-table show that if one uses these G-values,
eq. (\ref{eq.2}) holds quite well with the scatter rms of
$\left|G-G_{\rm obs}\right|/G_{\rm obs}=\left|\sigma_{\rm obs}-\sigma\right|/\sigma=23{\%}$.

For the remaining 26 metals, we need to inspect in detail,
primarily because the periodic table itself is rather complicated. 
There seem two ways of guessing $G$. In the first method, given the
observed $G_{\rm obs}$, we force to choose configurations counted from the
highest term until the sum of electron numbers becomes closest to $G_{\rm obs}$,
that is we round off $G_{\rm obs}$ to integer such that
$\left|G-G_{\rm obs}\right|\leq 0.5$, namely $G \equiv
(G_{\rm obs})_{\rm round}$. Naturally $G/G_{\rm obs}$ becomes almost unity as seen in
Fig. \ref{f1}(b)-below. Though there seems no reason to reject this first method,
we `feel uneasy' because Ag, K$^{19}$ and Ca$^{20}$, presumably simpler
than other metals, depart more from unity than other more complicated
metals, but also more importantly because many metals show much smaller deviations from
unity than the relative differences of `non-identical experimental
$\sigma$-values' between the K-P-table and PSJ-table ($\pm 7{\%}$ for 38
metals).

We adopt then an alternative second method in this paper as shown below. We add deeper
`electron configurations' (hereafter E-config) for some elements than the K-P-table; examples are
Li$^{3}$(from 2s$^{1}$ to $\underline{1s^{2}}$2s$^{1}$, leading to
$G=3$), V$^{23}$(3d$^{3}$4s$^{2} \rightarrow
$\underline{3p}$^{6}$3d$^{3}$4s$^{2}\rightarrow G=11$), and
Bi$^{83}$(6s$^{2}$6p$^{3}\rightarrow
$\underline{5s$^{2}$5p$^{6}$5d$^{10}$}6s$^{2}$6p$^{3}\rightarrow 23$).
Here the added part is underlined. The last one Bi, a typical semimetal, may be noteworthy, because by including enough deep
levels, it can be treated in the same way as others, and Bi gives
$\sigma \approx \sigma_{\rm obs}$ using $G_{\rm obs}=23.8$
(full E-config of Bi$^{83}$ is $[\{$Pd$^{46}\}$4f$^{14}]$5s$^{2}$5p$^{6}$5d$^{10}$6s$^{2}$6p$^{3})$. 
We could have assigned $G=21$ by excluding the first 5s$^2$, indicating non unique $G$-values for a large $G_{\rm obs}$.  
Another semimetal Sb$^{51}$($G$=10) discussed later can be treated similarly. Adding Y$^{39}$(G=11), Zr$^{40}$(G=12), La$^{57}$(G=11), 
and Hf$^{72}$(G=10 from 5p$^{6}$5d$^{2}$6s$^{2}$), altogether 7 metals fall in this category. 

Though it is possible to treat Sc, Ti and iron group similarly, we
introduce two rules below to obtain `much better' agreements with the
observations. We first introduce what we call (10$-x$)-rule. We examine
Fe$^{26}$(listed as 3d$^{6}$4s$^{2}$ in the K-P-table),
Co$^{27}$(3d$^{7}$4s$^{2})$ and Ni$^{28}$(3d$^{8}$4s$^{2})$, giving
tentative values of $G_{\rm tent}=8$, 9 and 10, respectively. However if we
introduce one rule that if $G_{\rm tent}>5$ is encountered in d-orbit where
the saturation is 10, we use $G=10-G_{\rm tent}$ as a subset of
d-orbit. Then $G/G_{\rm obs}$ becomes closer to unity. Namely,
$G_{\rm obs}$=(6.1, 3.8, 4.7) and new $G=$(2+4, 2+3, 2+2) are obtained for Fe, Co, and Ni, 
respectively, where the first numeral 2's come from s$^{2}$.
The result is $G/G_{\rm obs}=\sigma_{\rm obs}/\sigma$=(6/6.1, 5/3.8, 4/4.7)=(0.98, 1.3, 0.85).
On the other hand for
$G_{\rm tent}$ we would have obtained $G_{\rm tent}/G_{\rm obs}$=(1.6, 2.9, 2.6),
which we regard unsatisfactory. We applied this rule altogether to 6
elements in $8^{\rm th}-10^{\rm th}$ group, including Ru$^{44}$, Rh$^{45}$,
and Os$^{76}$.

A rule of this kind is seen in atomic
spectroscopy\cite{condon,allen}, where d$^{x}$ and d$^{10-x}$ give the
same LS coupling terms such as $^{1}$S,$^{1}$D,$^{1}$G, together with
similar rules like p$^{6-x}$ and f$^{14-x}$. Also when the cohesive
energy of many metals was estimated, a similar kind of rule has been
utilized\cite{friedel}, which is broadly consistent with an extensive
calculation\cite{moruzzi}. The only strong reason however that we use
(10$-x$)-rule is because it gives better agreements with the
observations.

\begin{figure}
\begin{center}
\includegraphics[width=8.5cm]{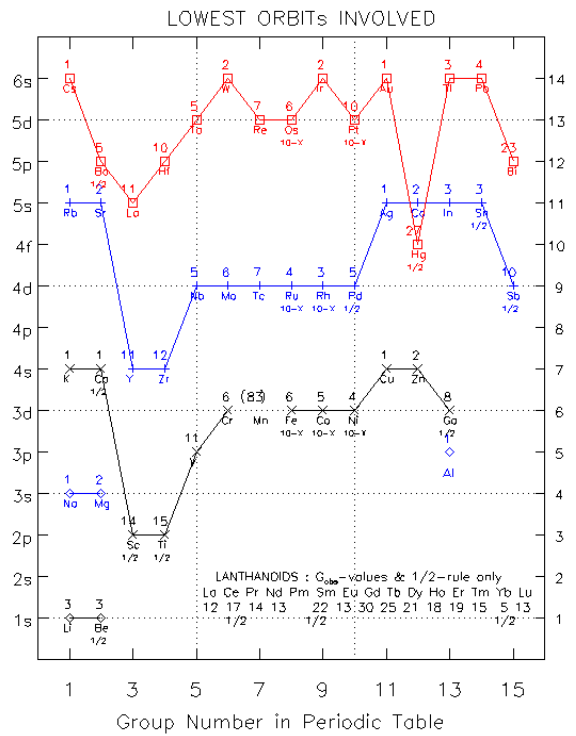}
\end{center}
\caption{The lowest positions in the electron configuration where the
 final count is made to fix $G$ from upper levels vs. group
 number. Numbers above metal names are $G$-values adopted (same as in
 Fig. \ref{f1}) and below names are whether (10$-x$) or (1/2) rule is
 employed.}
\label{f2}
\end{figure}

Finally we introduce what we call (1/2)-rule. When the $G_{\rm obs}$-value in a 
metal corresponds just inside of semi-closed shells of $g$=2, 6 or 10 (statistical weight), we
take one half of these values to the last of the sum of E-config. This is a sort of extended $(10-x)$-rule. 
Underlying presumption is that
though these 2, 6 and 10 electrons are closely packed, there might be
weak breaks just in the middle of these, namely 1, 3 and 5. As a first
example, Pd$^{46}$(4d$^{10}$) shows $G_{\rm obs}$=5.2 for
$\sigma$, hence instead of adopting $G=10$, we adopt $G=5$ from
10/2. We applied the `1/2-rule' also for Ca$^{20}$(4s$^{2}, G$=1, $G_{\rm obs}$=0.6), and Sn$^{50}$. 
For 6 metals we combine the addition of deeper terms to the K-P-table and the `1/2-rule';
Sc$^{21}$(\underline{2p$^{6}$3s$^{2}${3p}$^{6}$}3d$^{1}$4s$^{2}\rightarrow
G=14$), Ti$^{22}$(\underline{2p$^{6}$3s$^{2}$3p$^{6}$}3d$^{2}$4s$^{2}\rightarrow
G=15$) and Ba$^{56}$(\underline{5p$^{6}$}6s$^{2}\rightarrow G=5$). Here
added terms are underlined, and 6 in 2p$^{6}$-orbits (Sc and Ti) or
5p$^{6}$-orbits (Ba) is replaced by 6/2=3. Also Be$^{4}(G$=3), Ga$^{31}(G$=8) and Sb$^{51}(G$=10) fall in this group. 
Again it is noteworthy that though Sb is a semimetal, it is treated in the same way as the usual metal; 
namely $G$=10 for Sb$^{51}$, almost identical to $G_{\rm obs}$=10.0, comes from an addition of 5 
from K-P-table (5s$^{2}$5p$^{3}$), and 5=10/2 (1/2-rule for 3p$^{10}$)
from full E-config of Kr$^{36}$3p$^{10}$5s$^{2}$5p$^{3}$.
Further, the semimetal As$^{33}$ (Ar$^{18}$3d$^{10}$4s$^{2}$4p$^{3}$), not included so far, shows $G_{\rm obs}$=11.2 from $1/\sigma_{obs}$=333n$\Omega$m \cite{psj} so that
$G$=10 (1/2-rule for 3d$^{10}$) may be appropriate.  
Although use of the `(1/2)-rule' is due primarily to better fit the observations, we want to
stress that without this rule $\left|G-G_{\rm obs}\right| /G_{\rm obs}$ would
become much larger than in other metals in the same group where these
rules are not needed (see Fig. \ref{f2}).

Figure \ref{f1}(b)-upper shows, excluding problematic Mn$^{25}$, Mo$^{42}$, Pt$^{78}$, and
Hg$^{80}$, an rms scatter of $\pm 20$\% ($\pm 14$\% for $G=(G_{\rm obs})_{\rm round}$
in Fig. \ref{f1}(b)-lower), which is larger than the observation error of
(rms)$_{\rm obs}=\pm 7$\% mentioned before. This suggests that the scatter
in Fig. \ref{f1}(b)-upper stems largely from yet-unknown causes. 

For the thermal conductivity $\lambda$, we also show $G_{\rm obs-\lambda} \equiv \lambda_{1}/\lambda_{\rm obs}$ (Fig. \ref{f1}(a)) and
$\lambda/\lambda_{\rm obs}$ (Fig. \ref{f1}(b)-upper) 
from eq. (\ref{eq.2}) with the same $G$ used for $\sigma$ in thick marks
$[\lambda_{1} \equiv \lambda(G=1$)]. We plotted only
metals showing large departure from the Wiedemann-Franz law\cite{kittel,
ibach,aschcroft} ($\left|C_{\rm WF}-1 \right| \geq 0.2$). Here
$C_{\rm WF}\equiv\lambda/(\sigma TL_{\rm z})$, where $L_{\rm z}\equiv \pi^{2}k_{\rm B}^{2}/3e^{2}$. 
We find no appreciable differences from
$G_{\rm obs}=\sigma_{1}/\sigma_{\rm obs}$ even for those metals of large
$\left|C_{\rm WF}\!-\!1\right|$.

Figure \ref{f2}, which is supplementary to Fig. \ref{f1}, shows that the
position of estimated $G$'s in E-config. For example, E-config of
Fe$^{26}$ is \{Ar$^{18}$\}3d$^{6}$4s$^{2}$ and we adopted $G=2+4=6$ (4
comes from $10-6$ in d$^{6})$ added from outer ones, where 4 is within
the 3d-orbit. Hence for Fe we plotted at 3d. Fig. \ref{f2} shows rather
systematic behavior, particularly among 4-6 periods (starting
from K, Rb and Cs). This indicates that our choice of $G$, though
adopted only to match the observations, appears to be rooted from some
physical basis. In fact we notice that groups 3-4 (Sc$^{21}$,
Ti$^{22}$, Y$^{39}$, Zr$^{40}$, La$^{57})$, which stem deeper
configurations than other metals, are all have configurations where
electrons are filled in the outer orbits before inner orbits become
filled up or closed;
e.g. Sc$^{21}$=\{Be$^{4}$\}2p$^{6}$3s$^{2}$3p$^{6}$\underline{3d}$^{1}$\underline{4s}$^{2}$
instead of \{Be$^{4}$\}2p$^{6}$3s$^{2}$3p$^{6}$\underline{3d}$^{3}$,
while the 3d-orbit only saturates at Cu$^{29}$ as 3d$^{10}$. This is of
course typical characteristics of the earlier transition elements. It
might suggest some unstableness of so to speak \textit{heavier upper
floors} than e.g. noble metals, and as a consequence involvement of
deeper orbits. 

In addition to 48 metals plus As$^{33}$ already discussed, we show $G_{\rm obs}$-values for rare earth metals in Fig. \ref{f2} 
and find $\sigma / \sigma_{\rm obs}
\approx 1$, mainly because $G_{\rm obs}\ge10$ (except Yb$^{70}$) such that it is easier to find $G/G_{\rm obs}\approx 1$ 
(the lowest orbits are in 4d, 4f and 5p).
A conclusion from Fig. \ref{f2} is that our choice of $G$-values shows rather systematic
distribution among E-config for various metals, supporting the choice,
if not prove, besides giving nearly correct values of $\sigma_{\rm obs}$.

$\textit{Concluding Remarks.}$-- In our view, the reason why Cr($1/\sigma_{\rm obs} $
=129n$\Omega$m=12.9$\times10^{-6}\Omega$cm at $T_{\rm obs}$=295K ) 
is more resistive than Cu (17n$\Omega$m, 4s$^{1}, G$=1), in fact by 7.6, is simply due to large 
$G$(Cr, 3d$^{5}$4s$^{1}$)=6, i.e. a larger number of equally contributing orbits (or bands), 
since $n_{\rm atom}$(Cr)=8.3 is similar to 
$n_{\rm atom}$(Cu)=8.5 (10$^{28}$m$^{-3}$ unit).

We find in this paper that eq.(2) for $\sigma$ and $\lambda$ using $\tau=\hbar/k_{B}T$ [eq. (1)] agrees well with the observations for the majority of elemental metals, inclusive of semimetals, at room temperature ranges.
Here we adopt that $G$ is the sum of outer electron numbers in the electron configuration as listed in the back cover of Kittel's text with some modifications
(e.g. 3d$^{10}$4s is replaced by 4s in Cu).

Certainly one needs theoretical reasoning particularly on $\tau$ and we wish to know the value of $Z\equiv n_{\rm e}/n_{\rm atom}$, which
is not needed in this paper.
We emphasize again that $\tau_{\rm calc}/\tau_{\rm obs}=\sigma_{\rm calc}/\sigma_{\rm obs}$ is very close to unity
for alkali and noble metals where $G=1$ is adopted (T=295K); the ratios are e.g. 0.88(Na), 0.99(Rb), 1.24(Cs), 1.05(Cu), 
and 0.95(Au). The extreme simplicity of $\tau$ [eq.(1)] and good agreements with the experiments suggest that there might be
an `extremely' simple physical explanation for this, which we will discuss in the next paper along with the value of $Z$,
for which we will find $Z=1$ for $\sigma$ in majority of elemental metals.
\\

We thank  N. Miura and T. Suemoto for discussion. \\

\end{document}